\documentclass[twocolumn]{revtex4}

\usepackage{bm}%
\usepackage{graphicx}
\usepackage{setspace}
\usepackage[english]{babel}
\usepackage{amsmath}
\usepackage{SIunits}
\newcommand{\textcolor}[3][rgb]{#3}

\begin{document}
\newcommand{\InP}{\begin{otherlanguage}{english}InP:Fe\end{otherlanguage}~\xspace}

\title {\textbf{Temporal behavior of two-wave-mixing in photorefractive InP:Fe\\ versus temperature}}
\author{N.Khelfaoui}
\affiliation{Laboratoire Mat\'eriaux Optiques, Photonique et
Syst\`emes \\ Unit\'e de Recherche Commune \`a l'Universit\'e de
Metz et Sup\'elec - CNRS UMR 7132
 \\ 2, rue Edouard Belin, 57070
Metz Cedex, France}

\author{D.Wolfersberger}
\affiliation{Laboratoire Mat\'eriaux Optiques, Photonique et
Syst\`emes \\ Unit\'e de Recherche Commune \`a l'Universit\'e de
Metz et Sup\'elec - CNRS UMR 7132  \\ 2, rue Edouard Belin, 57070
Metz Cedex, France}
\author{G.Kugel}
\affiliation{Laboratoire Mat\'eriaux Optiques, Photonique et
Syst\`emes \\ Unit\'e de Recherche Commune \`a l'Universit\'e de
Metz et Sup\'elec - CNRS UMR 7132  \\ 2, rue Edouard Belin, 57070
Metz Cedex, France}
\author{N.Fressengeas}
\affiliation{Laboratoire Mat\'eriaux Optiques, Photonique et
Syst\`emes \\ Unit\'e de Recherche Commune \`a l'Universit\'e de
Metz et Sup\'elec - CNRS UMR 7132  \\ 2, rue Edouard Belin, 57070
Metz Cedex, France}
\author{M.Chauvet}
\affiliation{Institut FEMTO-ST Universit\'e de Franche Comt\'e
D\'epartement d'optique-UMR 6174 UFR Sciences et Techniques Route
de Gray 25030 Besan\c{c}on cedex, France }
\date{\today}

\begin{abstract}
 The  temporal response of two-wave-mixing in
 photorefractive InP:Fe
 under a dc electric  field at different temperatures has been studied.
In particular, the temperature dependence of the characteristic
time constant has been studied both theoretically and
experimentally, showing a strongly decreasing time constant with increasing temperature.
\end{abstract}
\maketitle

\section {Introduction}
The photorefractive effect leads to a variety of nonlinear optical
phenomena in certain types of crystals. The basic mechanism of the
effect is the excitation and redistribution of charge carriers
inside a crystal as a result of non-uniform illumination. The
redistributed charges give rise to a non-uniform internal electric
field and thus to spatial variations in the refractive index of
the crystal through the Pockels effect. Significant nonlinearity
can be induced by relatively weak ($\micro\watt$) laser radiation.
Phenomena such as self-focusing, energy coupling between two
coherent laser beams, self-pumped phase conjugation, chaos,
pattern formation and spatial soliton have attracted much
attention in the past 20 years \cite{Yeh93,Haw99}.

Among  photorefractive crystals, semiconductor materials  have
attractive properties for  applications in optical
telecommunications such as optical switching and routing. This is
 due to the fact that they are sensitive in the infrared region and their response time
can be fast ($\micro\second$)\cite{Sch02ol1229}.
\\
Two-wave-mixing is
an excellent tool to characterize the photorefractive effect in
these materials \cite{Idr87oc317, Pic89ap3798, Mar99ap77} by determining the gain of
amplification under  the influence of the applied field, impurity
densities, or grating period.
Some semiconductors, like InP:Fe, exhibit an intensity dependant
resonance at stabilized temperatures
\cite{Pic89ap3798,Idr87oc317}.

In this paper, we analyze the temperature dependance of
Two-Wave-Mixing (TWM) characteristic time constant,
theoretically at first and eventually against  experimental results. We propose
a formal description of the temporal evolution of carrier
densities in the medium, linking them to the TWM gain temporal
evolution.
\section {Time dependant space-charge field in $InP:Fe$}
The basic principles of the  photorefractive effect in InP:Fe are
well known \cite{Mar99ap77}. It involves three steps:
photoexcitation of trapped carriers into excited states, migration
of excited carriers preferentially towards non-illuminated regions
and capture into empty deep centers. This leads to the formation
of a local space-charge field $E_{sc}$ and thus to the modulation
of the refractive index. The modulated refraction index is then
able to interact  with  the beams that have
created it. When the modulation stems from beam interference as in two wave mixing, an
energy transfer between beams may occur.

The principle of  two-wave-mixing is to propagate simultaneously
in a photorefractive crystal two coherent beams,
which have an angle $\theta$ between their directions of
propagation. 
This phenomena is governed by the following system of coupled
nonlinear differential equations:
\begin{align}
\frac{d I_{s}}{d z}&=\frac{\Gamma.I_{s}.I_{p}}{I_{0}}-\alpha.I_{s}\\
 \frac{d
I_{p}}{dz}&=\frac{-\Gamma.I_{s}.I_{p}}{I_{0}}-\alpha.I_{p}
\end{align}
where $I_{p}$ is the pump intensity, $I_{s}$ is the signal
intensity, and $I_{0}$ is the total intensity equal to the sum
$I_{s}+I_{p}$, $\alpha$ is the absorption coefficient (assumed here to be the same for pump and signal). In a
photorefractive crystal, $\Gamma$ takes the following form
\cite{Pic89ap3798}:
\begin{eqnarray}
\Gamma_{0}&=&(\frac{2.\pi. n^{3}.
r_{eff}}{\lambda.\cos\theta}).Im\{E_{sc}\}    \label{gequa}
\end{eqnarray}

where $n$ is the refractive index, $r_{eff}$ is the effective
electro-optic coefficient, $\lambda$ is the beam wavelength in
vacuum
 and $Im\{E_{sc}\}$ is the  imaginary part of the space-charge
field $E_{sc}$(the $\frac{\pi}{2}$ shifted component of $E_{sc}$
with respect to the illumination grating). \textcolor[rgb]{0.98,0.00,0.00}{The expression of $E_{sc}$ will derived in the following lines}.
$\theta$ is the angle between the two beams.

In order to evaluate the photorefractive gain $\Gamma_{0}$ given
by equation (\ref{gequa}),  the space-charge field $E_{sc}$ has to
be calculated from the modified Kukhtarev model \cite{Kuk79fe949},
taking into account both electrons and holes as charge carriers.
We chose a model with one deep center donor, two types of carriers
(electrons and holes)\cite{str86ol312},
considering variations only in one transversal  dimension (x) as
described by the following set of equations:
\begin{subequations}\label{systeme}
 \begin{eqnarray}
\frac{d E}{d x}&=&\frac{e}{\epsilon}(N_{D}-N_{A}+p_{h}-n_{e}-n_{T})\label{systeme1}\\
 j_{n}&=&e\mu_{n}n_{e}E+\mu_{n}k_{b}T\frac{dn_{e}}{dx}\\
 j_{p}&=&e\mu_{p}p_{e}E-\mu_{p}k_{b}T\frac{dp_{h}}{dx}\\
 \frac{d n_{e}}{d t}&=&e_{n}n_{T}-c_{n}n_{e}p_{T}+\frac{1}{e}\frac{d j_{n}}{d x}\label{ndt}\\
  \frac{d p_{h}}{d t}
  &=&e_{p}p_{T}-c_{p}p_{h}n_{T}-\frac{1}{e}\frac{d j_{p}}{d x}\label{pdt}\\
 \frac{d n_{T}}{d t}&=&e_{p}p_{T}-e_{n}n_{T}-c_{p}p_{h}n_{T}+c_{n}n_{e}p_{T}\\
N_{T} &=&n_{T}+ p_{T}\\
 \int_{-d/2}^{d/2} Edx  &=&V_{app}  \label{systeme2}\\
E& =E_{app} +E_{sc}
\end{eqnarray}
\end{subequations}

where $E$ is the electric field, $n_{e}$ and $p_{h}$ are the
electron and hole densities in the respective conduction and
valence bands, $n_{T}=Fe^{2+}$ is the density of ionized occupied
traps, $p_{T}=Fe^{3+}$ is the density of neutral unoccupied traps,
$J_{n}$ and $J_{p}$ are respectively  the electron and hole
currents. $N_{T}$, $N_{D}$ and $N_{A}$ are respectively  the
densities of iron atoms, the shallow donors and the shallow
acceptors. The charge mobilities are given by $\mu_{n}$ for
electrons and $\mu_{p}$ for holes, the electron and hole
recombination rate are respectively  $c_{n}$ and $c_{p}$ , $T$ is
the temperature and $k_{b}$ is the Boltzmann constant. The
dielectric permittivity is given by $\epsilon$ while $e$ is the
charge of the electron. $V_{app}$ is the voltage  applied
externally to the  crystal of width $d$. The electron and hole
emission parameters are $e_{n}$ and $e_{p}$  depend on both thermal and optical emission as
described by:
\begin{align}
e_{n}&=e^{th}_{n}+\sigma_{n}\frac{I(x)}{h\nu}\label{elec}\\
e_{p}&=e^{th}_{p}+\sigma_{p}\frac{I(x)}{h\nu}\label{trou}
\end{align}
where the thermal contribution to the emission rate coefficient is
$e^{th}$ and the optical cross section of the carriers is given
by $\sigma$, $I(x)$ is the spatially dependent intensity of
light due to the interferences between pump and signal beams and $h\nu$ is the photon energy.

For sufficiently small modulation depth $m$, intensity and all carriers densities may be expanded into Fourier series interrupted after the first term :
\begin{equation}\label{fourier}
A(x)=A_{0}+ A_{1}e^{iK_{g}x}
\end{equation}
where $A(x)$ takes the role of $I$, $n_{e}$, $p_{h}$, $n_{T}$, $p_{T}$ and  $K_{g}$ the spatial
frequency of the interference pattern. So the light intensity can be written for the average intensity $I_{0}$ as:

\begin{equation}
 I(x)=I_{0}(1+me^{iK_{g}x})
\end{equation}

In  the following, we have calculated the temporal evolution of carriers density and we will look forward to finding the temporal evolution of the space charge field under these hypothesis, i.e. considering only the zero'th and first order of the Fourier expansion.

The zero'th order corresponds to an uniform illumination ( $I(x)=I_{0}$). The space charge field is thus equal to zero and the local field is uniform and equals the applied field $E_{app}$. The electrons and holes densities at steady state  are known to be equal to $\frac{e_{n}.n_{T_{0}}}{c_{n}.p_{T_{0}}}$ and $\frac{e_{p}.p_{T_{0}}}{c_{p}.n_{T_{0}}}$ respectively \cite{Pic89ap3798}, where $n_{T_{0}}$, $p_{T_{0}}$ are the density of occupied  and unoccupied traps at steady state.
\begin{figure}
\center
\includegraphics [width=8cm]{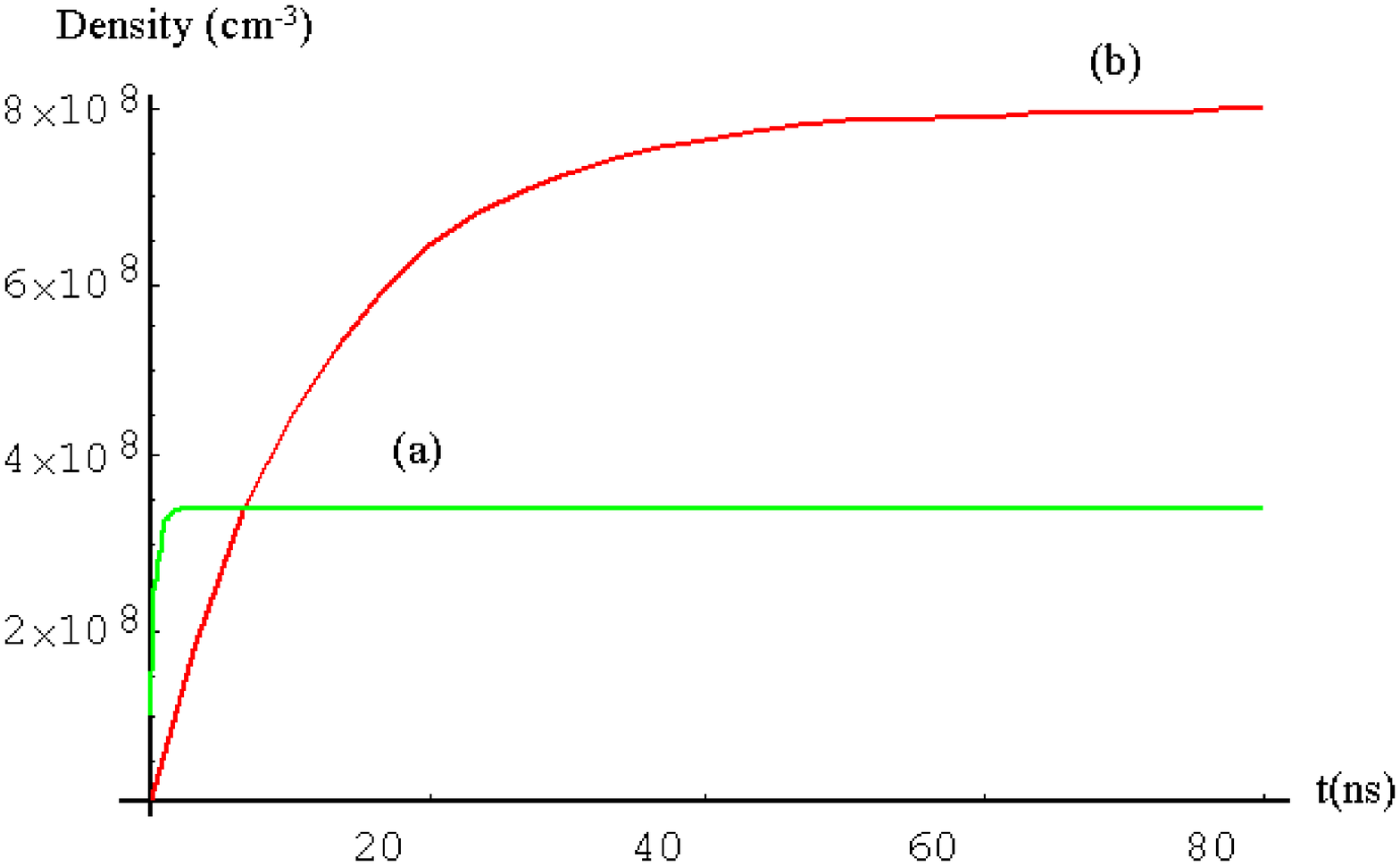}
\caption{ Temporal evolution of electron  (a) and  hole densities
(b) under uniform illumination for $\lambda = 1.06~ \micro\metre$,
$I_{0} = 20~\milli\watt\per\centi\metre\squared$, T=297K. Materials
parameters taken from ref \cite{Pic89ap3798} are:~$c_{n}
=
  4.1\times 10^{-8}\centi\metre^{3}\per\second$,~$c_{p} = 1.6\times10^{-8}\centi\metre^{3}\per\second$,~$e_{p}^{th}=10^{-4}\second^{-1}$,~$e_{n}^{th}=16.32\second^{-1}$,~ $n_{T_{0}} =5\times10^{15}\centi\metre^{-3}$,~$p_{T_{0}} =
  6\times10^{16}\centi\metre^{-3}$,~$n_{0}=1\times10^{7}\centi\metre^{-3}$,~$p_{0}=6\times10^{6}\centi\metre^{-3}$,~$\sigma_{n}= 4\times10^{-18}\centi\metre^{2}$,~$\sigma_{p} = 1\times10^{-17}\centi\metre^{2}$ .}
\label{ordrezeroelectron}
\end{figure}

The electrons and holes densities in transient regime when an uniform
illumination is established, are calculated by solving equations
(\ref{ndt}) and (\ref{pdt}) , assuming that $I_{0}=0$  for $t<0$ and at time $t=0$,
the carriers density values are equal to $n_{0}$ and $p_{0}$ at
thermal equilibrium, without any optical excitations \cite{Haw99}.
We obtained the following solution:

\begin{align}
n_{e}(t)&=\frac{e^{-c_{n}p_{T_{0}}t}(c_{n}p_{T_{0}}n_{0}+(-1+e^{c_{n}p_{T_{0}}t})n_{T_{0}}(e_{n}^{th}+\sigma_{n}\frac{I_{0}}{h\nu}))}{c_{n}p_{T_{0}}}\\
p_{h}(t)&=\frac{e^{-c_{p}n_{T_{0}}t}(c_{p}n_{T_{0}}p_{0}+(-1+e^{c_{p}n_{T_{0}}t})p_{T_{0}}(e_{p}^{th}+\sigma_{p}\frac{I_{0}}{h\nu}))}{c_{p}n_{T_{0}}}
\end{align}

The temporal evolution of carrier densities under uniform
illumination is illustrated in figure \ref{ordrezeroelectron}. Our
model confirms the fact that the carrier densities evolution grows
exponentially. The rise time of carriers generation is on the
order of nanosecond time scale for a beam intensity of a few
\milli\watt\ per \centi\metre\squared. Without  presence of the
beam, the electron density is greater than the hole density
because  electrons are mostly generated thermally while holes are
generated optically \cite{Pic89ap3798}.

For a modulated intensity (first Fourier order), by using the set
of equations (\ref{systeme}), the space-charge field can be \textcolor[rgb]{0.98,0.00,0.00}{approximatively }
expressed at steady state  as \cite{Pic89ap3798}:
\begin{equation}\label{Esta}
\textcolor[rgb]{0.98,0.00,0.00}{E_{1}=\frac{i.m.I_{0}}{(I_{res}+
I_{0})(\frac{1}{E_{q}}+\frac{E_{d}}{E_{0}^{2}+E_{d}^{2}})
+i(I_{res}-I_{0})\frac{E_{0}}{E_{0}^{2}+E_{d}^{2}}}\approx m.E_{sc}}
\end{equation}
\\
where $E_{0}$, $E_{d}=K_{g}\frac{k_{b}.T}{e}$ and $E_{q} =\frac{e}{\epsilon.
K_{g}}.\frac{n_{T_{0}}.p_{T_{0}}}{n_{T_{0}}+p_{T_{0}}}$ are the space charge field under uniform illumination, the diffusion field and charge-limiting field respectively. $I_{res}= \frac{e_{n}^{th}.n_{T_{0}}}{\sigma_{p}.p_{T_{0}}.h.\nu}$ is the
resonance intensity  defined as the intensity at
which holes and electrons are generated at the same rate.

From equation (\ref{Esta}), we observe that  the space charge
field is purely imaginary, when the illumination $I_{0}$ equals
$I_{res}$. Above resonance, the hole density is higher than the electron one, mainly because the holes cross section is stronger
than for electrons.
The result is that charge transfer mainly occurs between iron
level and the valence band. Below resonance, when electrons are
dominant, the iron mainly interacts with the electrons and the
conduction band.
\\In transient regime for a  modulated intensity, the dynamics of the space charge field
is calculated by considering an adiabatic approximations
\cite{valqe1983}, a concentration of electrons and holes densities
reaches instantaneously the equilibrium value which depends on the
actual concentration of filled and empty traps, so we set:
$\frac{dp_{h}}{dt}=\frac{dn_{e}}{dt}=0$. We assume that the
electrons are excited thermally while the holes optically
\cite{Pic89ap3798}. In the low modulation approximation, some
algebraic manipulations of the set of  equations (\ref{systeme})
lead to:

\begin{equation}
E_{1}(t) = m E_{sc}[1-exp(-\frac{t}{\tau_{g}})]\label{e1temp}
\end{equation}

where $\tau_{g}$ is a  complex time constant, which can be
rewritten by separating its real and imaginary parts.
\begin{align}\label{hgf}
    \frac{1}{\tau_{g}}&=\frac{1}{\tau}+iw
    \end{align}
\textcolor[rgb]{0.98,0.00,0.00}{with}  \begin{align}
 \tau&=\frac{\tau_{n}\tau_{p}}{\tau_{n}+\tau_{p}}\notag
 \end{align}
\textcolor[rgb]{0.98,0.00,0.00}{and}  \begin{align}
 w&=w_{n}-w_{p}\notag
\end{align}

\textcolor[rgb]{0.98,0.00,0.00}{The subscript indexes $n$ and $p$ are related to the electron and hole contributions respectively. $\tau_{n}$ and $w_{n}$ are given by:}
\begin{align}\label{to}
\tau_{n}&=\tau_{di,n}\frac{(1+\frac{Ed}{E_{Mn}})^{2}+(\frac{E_{0}}{E_{Mn}})^{2}}{(1+\frac{Ed}{E_{Mn}})(1+\frac{Ed}{Eq})+\frac{E_{0}^{2}}{(E_{Mn}.Eq)}}
\end{align}

\begin{align}\label{www}
w_{n}&=\frac{1}{\tau_{di,n}}\frac{\frac{E_{0}}{E_{Mn}}-\frac{E_{0}}{Eq}}{(1+\frac{Ed}{E_{Mn}})^{2}+(\frac{E_{0}}{E_{Mn}})^{2}}
\end{align}
\textcolor[rgb]{0.98,0.00,0.00}{where $E_{Mn}=\frac{c_{n}.p_{T_{0}}}{\mu_{n}.k_{b}}$ is the mobility field,
$\tau_{di,n}$ is the electron dielectric relaxation time depending on intensity and temperature which can be written as:
\begin{equation}\label{dodo}
   \tau_{di,n}=\frac{e_{n}^{th}.n_{T_{0}}.e.\mu_{n}}{c_{n}.p_{T_{0}}.\epsilon}
\end{equation}
$e_{n}^{th}$ is the thermal parameter equal to:
\begin{equation}\label{toutou}
    e_{n}^{th}=3,25.10^{25}.\frac{m_{n}^{*}}{m}.T^{2}.\sigma_{n}^{\infty}.e^{\frac{-E_{na}}{k_{b}.T}}
\end{equation}
where $\frac{m_{n}^{*}}{m}$ is the effective masse of electron, $E_{na}$ is the apparent activation energy of the electron trap,  $\sigma_{n}^{\infty}$ is the electron capture cross section. The value of this parameters are determined experimentally \cite{bre81el55}.\\
 To obtain $\tau_{p}$, $w_{p}$ and $\tau_{di,p}$, the index $p$ should be substituted for the index $n$ in the equations (\ref{to}), (\ref{www}) and (\ref{dodo}).\\
From equations (\ref{e1temp}) to (\ref{toutou}), it is possible to deduce, as was done previously \cite{Idr87oc317, Pic89ap3798}, that the time constant $\tau_{g}$ is real if electron emission is equal to the hole emission. That is, in the case of InP:Fe, electron thermal-emission is equal to the holes optical-emission. This allows to infer a link between the behavior of InP:Fe as a function of intensity and temperature.\\It will be the aim of the following sections to confirm this link, both theoretically and experimentally}

\section{Gain dynamics}
The photorefractive gain $\Gamma$ is the main parameter that can be determined by
  two-wave-mixing. It quantifies  an
energy transfer from the pump beam to the  signal beam and  is
proportional to the imaginary part of the space-charge field.

The gain value depends on different parameters like applied
electric field, iron density $N_{T}$, pump intensity and
temperature \cite{Ozksab1997}. Our work concentrate on the study
of  the gain dynamics versus temperature and we particularly
analyze the dependance of the rise time on temperature.

The stationary value of the photorefractive gain at different
temperatures is given by equation (\ref{gequa}) where $E_{sc}$ \textcolor[rgb]{0.98,0.00,0.00}{is
given by equation (\ref{Esta})}. This expression
shows that a maximum  gain is reached when $I_{0}=I_{res}$. This
maximum corresponds to an intensity resonance \cite{Pic89ap3798}.

We studied theoretically the temporal gain behavior using the
standard definition given by equation (\ref{gaist}) deduced from
equation (\ref{e1temp}) by developing $E_{sc}$ and $\tau_{g}$.

\begin{equation}
\Gamma=\Gamma_{0}[1+\exp(\frac{-t}{\tau})\times\frac{\sin(wt-\psi)}{\sin
\psi}]\label{gaist}
\end{equation}

where\\$\psi$: argument of stationary space-charge field
($E_{sc}=|E_{sc}|\exp{i\psi}$)
\\$\tau$: amplification's characteristic rise time.

\begin{figure}
\center
\includegraphics [width=8cm]{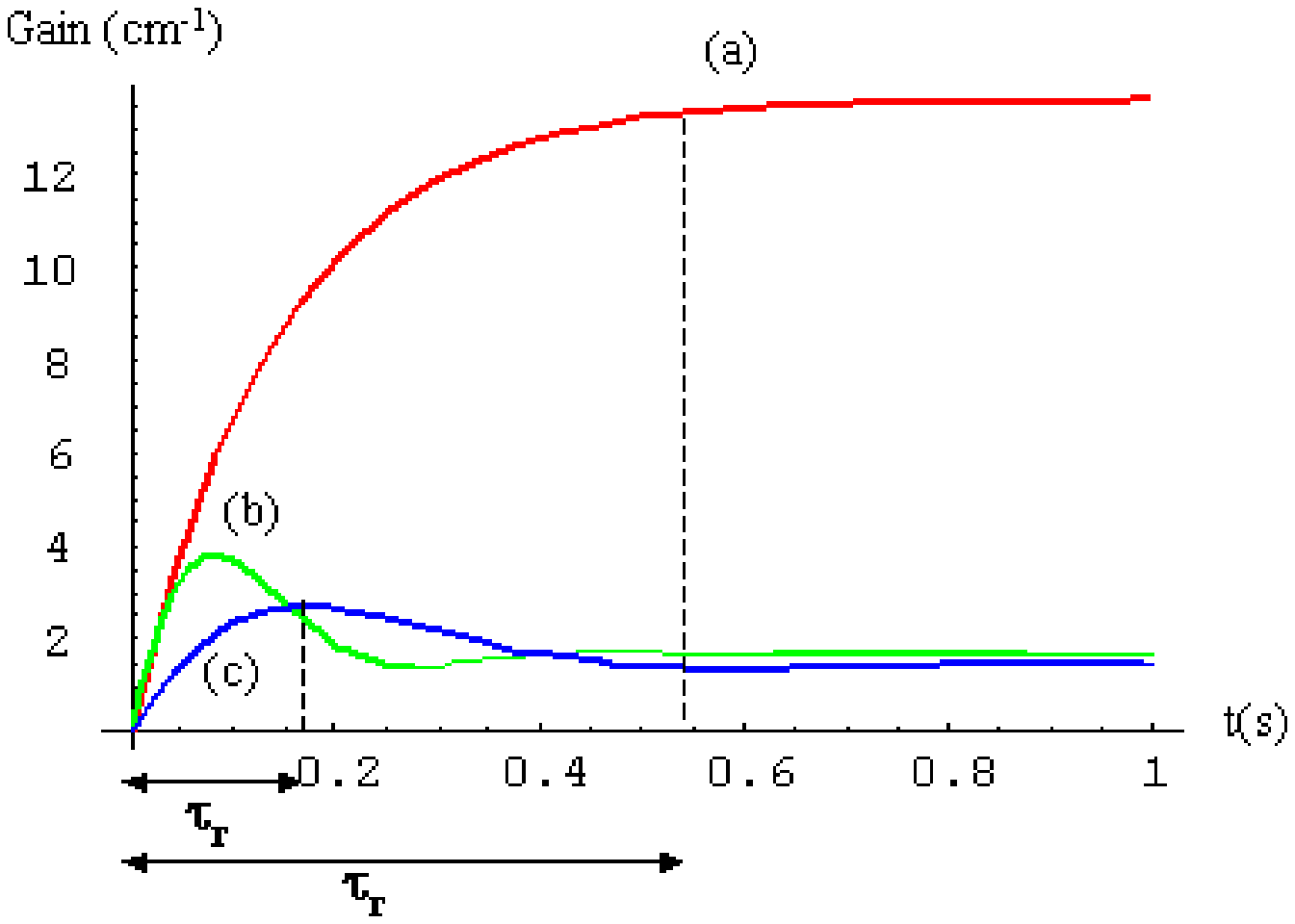}
\caption {Temporal evolution of local gain at  different  intensities at
$T = 297 K $ \textcolor[rgb]{0.98,0.00,0.00}{($e_{n}^{th}=16.31\second^{-1}$ calculated from equation \ref{toutou})},~~$\Lambda = 5~ \micro\metre$,~~$E_{0}
=10~\kilo\volt\per\centi\metre$, crystal thickness $L=12mm $ : (a)
$I_{0}=25.5~\milli\watt\per\centi\metre\squared \sim Ires$, (b)
$I_{0}=15~\milli\watt\per\centi\metre\squared$,~~(c)$I_{0}=50~\milli\watt\per\centi\metre\squared$. ~$\tau _{r}$: characteristic
time constant of amplification.}
\label{gaintempintes}
\end{figure}

Our theoretical simulations produce the curves represented on
figure \ref{gaintempintes},  illustrating the evolution of
photorefractive gain as function of time for three different pump
beam intensities: at resonance, below and above resonance for
 the same parameters as in figure \ref{ordrezeroelectron}. We see that  the gain amplitude  differs from each intensity to another, it takes the maximum value around resonance.

\begin{figure}
\center
\includegraphics [width=8cm]{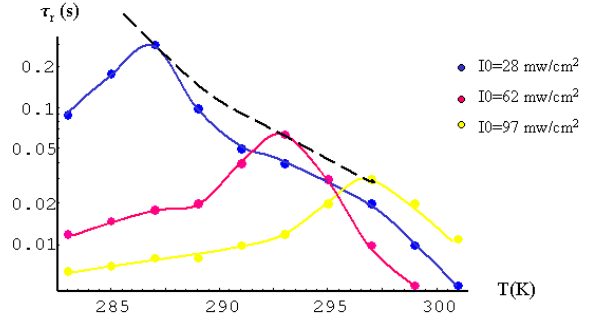}
\caption {\textcolor[rgb]{0.98,0.00,0.00}{Characteristic time of amplification versus temperature
at three different intensities. The time constant at the resonance intensity is given for each temperature by the dotted line. }}
\label{tempeires}
\end{figure}

As a next step, we studied  theoretically the TWM gain time
response as a function of temperature.
For an easier comparison with experimental results, in the
following, the response time is $\tau _{r}$ will be considered 
\textcolor[rgb]{0.98,0.00,0.00}{as the time interval
necessary for the gain to reach $90\%$ of the first maximum of each curves} as shown in figure \ref{gaintempintes}. The response time $\tau _{r}$ versus
temperature are given in figure \ref{tempeires} for three
distinct intensities, along with a fourth fitted curve showing the time
constant at resonance intensities.
We observe that the response time quickly decays as temperature
increases. At resonance, $\tau _{r}$  is larger because the space
charge field is high and it consequently necessitates more charge
to accumulate.

\section{Average gain}
The theoretical curve shown in figure \ref{gaintempintes}, illustrates the temporal evolution of the local gain. For the InP:Fe sample, the absorption coefficient at $\lambda=1.06 \mu m$ being approximately equal to $1 cm^{-1}$. Owing to this absorption, the mean intensity decrease along the $z$ axis  propagation.
The exponential gain would result from an integration over the optical thickness, as described in equation \ref{average}.
\begin{equation}
\Gamma=\frac{1}{L}\int^{L}_{0}\Gamma(z).dz \label{average}
\end{equation}

\begin{figure}
\center
\includegraphics [width=8cm]{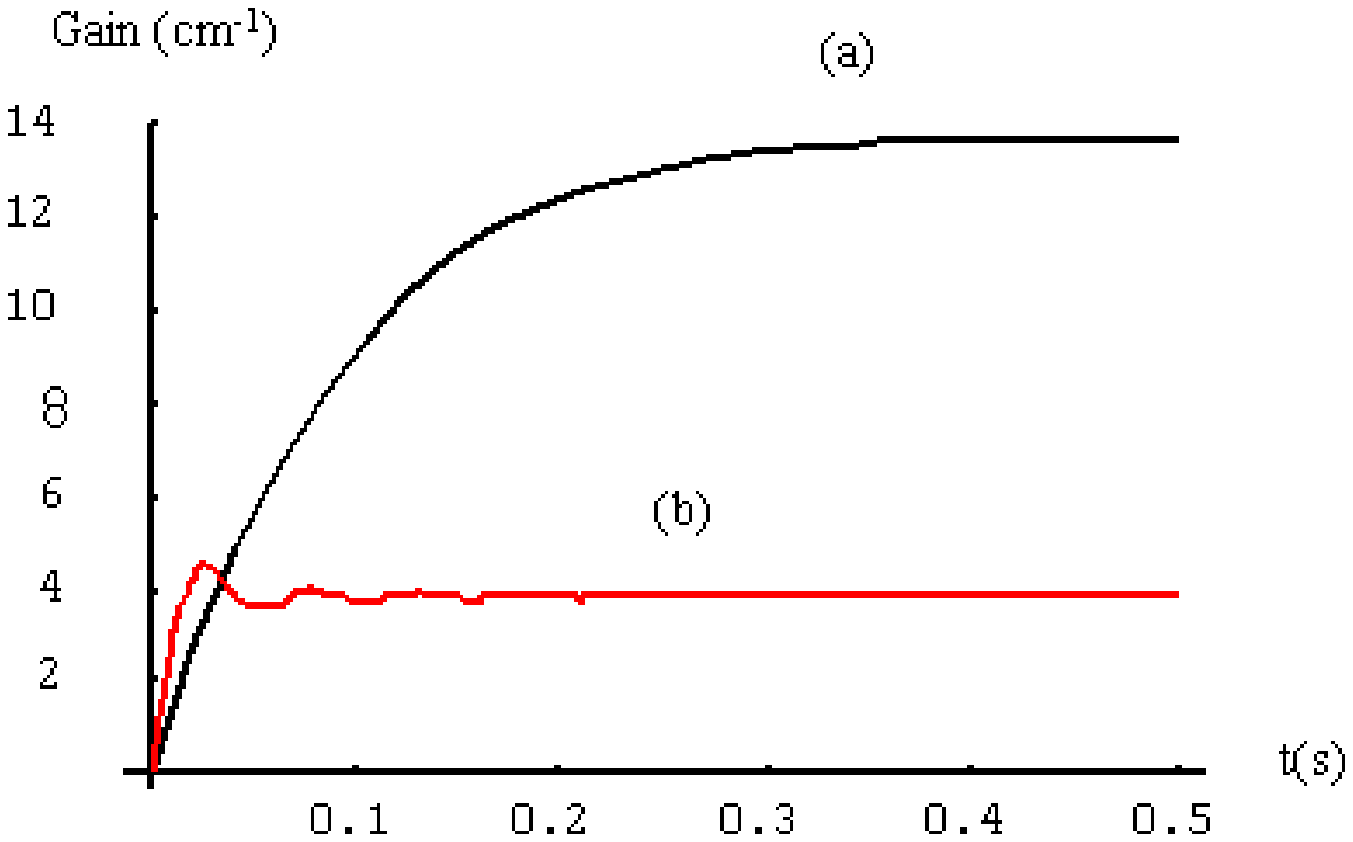}
\caption {Local (a)  and average gain (b) versus time,~$I_{0}=25.5~\milli\watt\per\centi\metre\squared$,~
$e_{n}^{th}=16.31\second^{-1}$,~$\Lambda = 5~ \micro\metre$,~$E_{0}
=10~\kilo\volt\per\centi\metre$.}
\label{Compa}
\end{figure}

\begin{figure}
\center
\includegraphics [width=8cm]{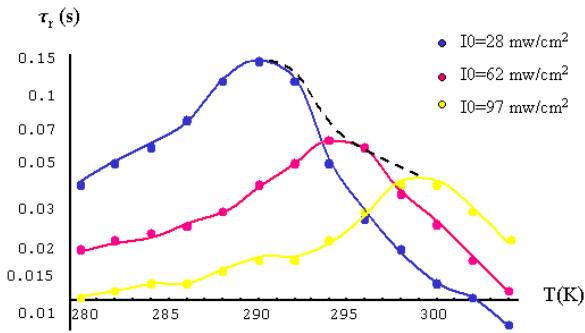}
\caption {Average gain characteristic time  versus temperature
at three different intensities. The time constant at the resonance intensity is given for each temperature by the dotted line. }
\label{tempeires1}
\end{figure}
The figure \ref{Compa} shows temporal evolution of local and average gain for $L=12mm$ crystal thickness  for the same intensity; the average gain is lower because the intensity absorption is taken into account.

Because of the absorption the resonance intensity for average gain
is higher than the local one for the same temperature. As for the
local gain, we calculated numerically the characteristic time, in
the same way.
 The results are shown in figure \ref{tempeires1}.
 We  compare the results in figure \ref{tempeires}, we observe the following  differences: the resonance
  peaks are slightly widened because is reached  within the example for various input intensities
   and the peaks  are shifted towards high intensities again because of absorption.
   These conclusions can arise from figures \ref{tempeires} and \ref{tempeires1}
    although they show the rise time as a function of temperature.
     Indeed, our calculations  show that the photorefractive gain and rise time are linked,
      so that the rise time is the slowest for the highest gain (i.e. at
      resonance); since more charges need to be accumulated.

\section{Experimental validation}
\begin{figure}
\center
\includegraphics [width=8cm]{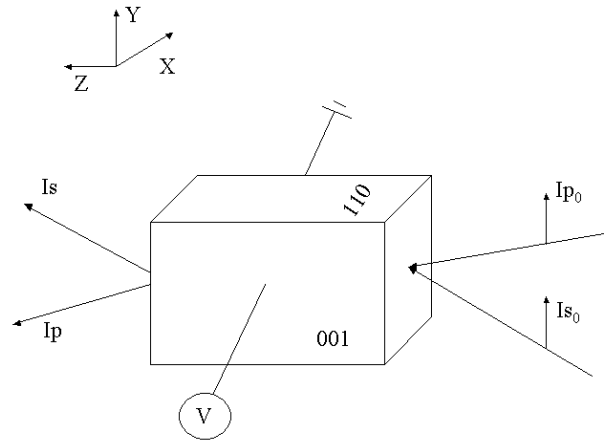}
\caption{Two-wave-mixing  configuration. } \label{coupling}
\end{figure}

We perform standard two-wave mixing experiments in co-directional
configuration as shown in figure \ref{coupling}. Pump and signal
beam intensities  ratio is set to $\beta =50$ and the angle
between pump and signal is $2\theta=12^{0}$ corresponding for an
space grating $\Lambda
 =5\mu m$. The experiments are performed with a CW
$1.06\micro\metre$ YAG laser.

An electric field ($10 \kilo\volt\per\centi\metre$) is applied
between the $<001>$ faces of InP:Fe crystal ($5\times 5\times12
mm^{3}$). The light beam is linearly polarized along the
$<\overline{1}10>$ direction and propagates along the $<110>$
direction ($12mm$). The absorption constant as measured by
spectrometer is close to $1 cm^{-1}$ at $1.06 \mu m$. Crystal
temperature is stabilized by a Peltier cooler.

\begin{figure}
\center
\includegraphics [width=8cm]{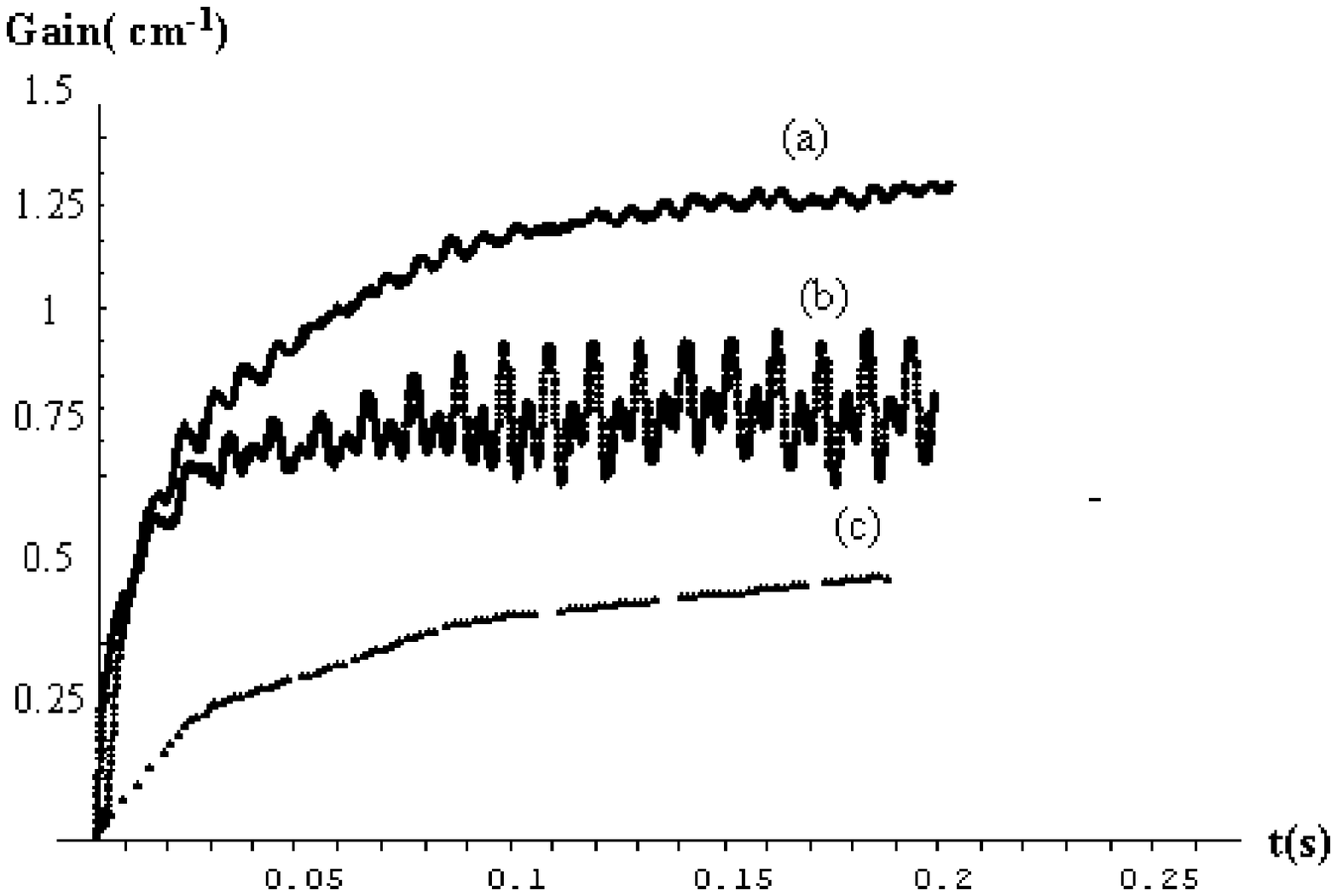}
\caption {Gain dynamics at 288K at different
intensities: (a)$62~\milli\watt\per\centi\metre\squared$,(b)$97~\milli\watt\per\centi\metre\squared$
(c)$28~\milli\watt\per\centi\metre\squared$.} \label{curve}
\end{figure}

\begin{figure}
\center
\includegraphics [width=8cm]{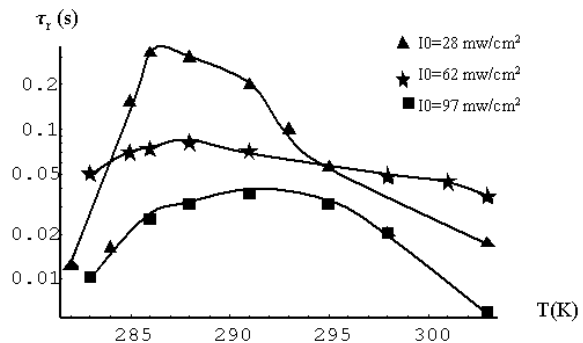}
\caption {The $\tau _{r}$  versus temperature at different
intensities.} \label{nouveaupoints}
\end{figure}

Transient behavior is analyzed by measuring $\tau _{r}$ as was done in figures \ref{tempeires} and \ref{tempeires1}. Figure \ref{curve}
shows the results obtained for three different intensities from
one side to the other of the resonance (the oscillations seen on figure \ref{curve} are attributed to the experimental noise and the curves are assumed to correspond to the first order responses).
Experimental results concerning the TWM time constant are given on
figure \ref{nouveaupoints}. For high temperature, $\tau _{r}$
decreases for all intensities.

Note that, for both theory and experiments, $\tau _{r}$  value
decreases from $300$ to $50~\milli\second$ for an increase
temperature of $10^{0}C$--showing a good quantitative result. The
discrepancy observed between figure \ref{nouveaupoints} and
\ref{tempeires} is partially corrected by taking into account  the
gain integration along the beam path inside the crystal, as shown
in figure \ref{tempeires1}, showing a widening of the curves. We
attribute the difference observed in terms of gain maximum value
to lack of precision in the knowledge of the crystal's physical
constants such as photo-excitation, cross section  and dopant
concentration.

\section{Conclusion}
We have studied the dynamics of TWM in InP:Fe as a function of
intensity and temperature. A theoretical analysis shows that the
gain coefficient oscillates when an intensity lower or higher than
the resonance intensity is used. At resonance the gain grows
exponentially.

The experimental study shows that the crystal absorption prevents
 the oscillating behavior. We have shown that the
gain rise time is strongly temperature dependent. Experimentally
the gain rise time is  $10$ times shorter at $295 K$ than at $285
K$ for low intensities.  

According to experimental and applications needs, the temperature
as well as the intensity can be used to tune the photorefractive
response time.


\end{document}